\documentclass[sigconf,
]{acmart}

\pdfoutput=1

\usepackage{booktabs} % For formal tables

\setcopyright{none}
\acmConference[WSDM]{The First International Workshop on Context-Aware Recommendation Systems with Big Data Analytics (CARS-BDA), co-organized with the 12th ACM International Conference on Web Search and Data Mining}{2019}{Melbourne, Australia}
\acmYear{2019}
\copyrightyear{2019}

\begin{document}
%Utilizing Negative Feedback for Improving Recommendation Quality
\title[Utilizing Negative User Preference to Improve Recommendation Quality
%in Collaborative Metric Learning
]{Loss Aversion in Recommender Systems:
\\Utilizing Negative User Preference to Improve \\ Recommendation Quality
%Improving Recommendation Quality with \\Two-Class Collaborative Metric Learning
}
%\titlenote{Produces the permission block, and
%  copyright information}
%\subtitle{Extended Abstract}
%\subtitlenote{The full version of the author's guide is available as
%  \texttt{acmart.pdf} document}

\author{Bibek Paudel}
%\authornote{}
\affiliation{%
  \institution{Department of Informatics\\University of Z\"urich}
  \city{Z\"urich}
  \state{Switzerland}
}
\email{paudel@ifi.uzh.ch}

\author{Sandro Luck}
\affiliation{%
  \institution{University of Z\"urich}
  \city{Z\"urich}
  \state{Switzerland}
}
\email{sandro.luck@uzh.ch}

\author{Abraham Bernstein}
\affiliation{%
  \institution{Department of Informatics\\University of Z\"urich}
  \city{Z\"urich}
  \state{Switzerland}
  }
\email{bernstein@ifi.uzh.ch}

\renewcommand{\shortauthors}{B. Paudel et al.}

\begin{abstract}
Negative user preference is an important context that is not sufficiently utilized by many existing recommender systems. This context is especially useful in scenarios where the cost of negative items is high for the users. 
In this work, we describe a new recommender algorithm that explicitly models negative user preferences in order to recommend more positive items %while reducing the amount of negative items
at the top of recommendation-lists. 
We build upon existing machine-learning model to incorporate the contextual information provided by negative user preference. With experimental evaluations on two openly available datasets, we show that our method is able to improve recommendation quality: by improving accuracy and at the same time reducing the number of negative items at the top of recommendation-lists. Our work demonstrates the value of the contextual information provided by negative feedback, and can also be extended to signed social networks and link prediction in other networks.
\end{abstract}


\begin{comment}
\begin{CCSXML}
<ccs2012>
 <concept>
  <concept_id>10010520.10010553.10010562</concept_id>
  <concept_desc>Computer systems organization~Embedded systems</concept_desc>
  <concept_significance>500</concept_significance>
 </concept>
 <concept>
  <concept_id>10010520.10010575.10010755</concept_id>
  <concept_desc>Computer systems organization~Redundancy</concept_desc>
  <concept_significance>300</concept_significance>
 </concept>
 <concept>
  <concept_id>10010520.10010553.10010554</concept_id>
  <concept_desc>Computer systems organization~Robotics</concept_desc>
  <concept_significance>100</concept_significance>
 </concept>
 <concept>
  <concept_id>10003033.10003083.10003095</concept_id>
  <concept_desc>Networks~Network reliability</concept_desc>
  <concept_significance>100</concept_significance>
 </concept>
</ccs2012>
\end{CCSXML}

\ccsdesc[500]{Computer systems organization~Embedded systems}
\ccsdesc[300]{Computer systems organization~Redundancy}
\ccsdesc{Computer systems organization~Robotics}
\ccsdesc[100]{Networks~Network reliability}
\end{comment}

\keywords{recommender systems, collaborative filtering, metric learning}

\maketitle

\section{Introduction}
%\vspace{-1mm}
Recommender systems are used in a variety of fields like online-shopping, music streaming and movie rental services. They have a big impact on how items are perceived by and displayed to users. These systems offer users many advantages such as decreasing the search time and improving user satisfaction.
Several machine learning algorithms have been developed and researched to improve the quality of suggestions generated by recommender systems. 

While most existing recommender systems model only positive feedback, in this work we focus on the negative feedback given by users. By incorporating the context provided by negative feedback, we propose a novel machine learning algorithm which is able to improve the quality of recommendations.

Our work is motivated by a well-known theory in behavioral economics and cognitive psychology.
According to the \emph{loss aversion theory}~\cite{kahneman2013prospect}, people value loss and gain differently --- the pain from a loss is psychologically more powerful than pleasure from a similar gain.
For instance, the feeling of disappointment over a poor book recommendation can outweigh the satisfaction provided by a good book suggestion.
%The effect of a negative recommendation and resulting feeling of loss can be perceived more by users in some areas (e.g., a bad job recommendation) compared to others (e.g., a bad song recommendation).
%
Therefore, to improve user experience, it is important for recommender systems to make a distinction between positive and negative user-feedback. 
The goal should not just be to recommend positive items, but also to suggest fewer negative items to users.
%Recommending fewer negative items is likely to improve user experience than only focusing on recommending positive items.

We build upon previous work from the domains of recommender systems and deep feature learning. 
Specifically, we utilize recent work on Collaborative Metric Learning (CML)~\cite{hsieh2017collaborative} and Two-Class Collaborative Filtering (TCCF)~\cite{paudel2017fewer} that have been shown to produce more diverse and accurate recommendations. 
Both these models are based on Collaborative filtering (CF), which is a popular approach in recommender systems. CF is a process where information about similar users (or items) is collaboratively used to filter information. The goal is to predict user interactions on different items (e.g. recommend list of movies the user might like).

Most existing recommender systems produce personalized rankings for users by differentiating previous positive choices from all other choices. 
They treat user preference as a binary variable: known positive preferences are considered positive, and unknown preferences are considered negative. 
These algorithms only optimize the benefit of putting positive items at the top of the recommendations, and ignore the cost of negative items as they are considered similar to unknown items. 
In several scenarios, it is undesirable for users to have negative items at the top of their recommendations. 
This motivates the problem of Two-Class Collaborative Filtering, where the goal is to recommend fewer negative items at the top while maintaining high prediction accuracy.

In many cases and datasets, explicit or implicit negative feedback from users are available, making the need to explicitly model them more timely. Likes/dislikes in online platforms, friend/foe relations and signed social networks are some examples where explicit negative feedback is common. In some scenarios, implicit negative feedback like low-ratings (e.g., ratings of 1-2 in a scale of 1-5) are available. In this work, we focus on the recommendation problem, but our work can be applied in other problems, like link prediction and modeling of signed networks.

Our contributions in this paper are as follows: (i) we present Two-Class Collaborative Metric Learning (TC-CML), a novel recommendation algorithm based on CML, in which we explicitly model negative feedback, (ii) we evaluate our approach on two benchmark datasets and and find that compared to CML, our approach is able to improve recommendation accuracy as well as reduce the number of negative items at the top of recommendations. In the remainder of this paper, we describe previous work, present our model and the results of experimental evaluation.
%\vspace{-2mm}
\section{Related Work}
Metric learning (ML)~\cite{xing2003distance} algorithms learn distance metrics that capture relations between items.
They learn a metric that assigns a low distance to similar items and a high distance to dissimilar items.

One recent method developed for recommender systems, called Collaborative Metric Learning (CML)~\cite{hsieh2017collaborative}, uses metric learning to embed users and items in a low-dimensional vector space.
The embeddings can then be used to rank items which a user is likely to enjoy.
It has been shown to generate better recommendations than other state-of-the-art systems.
CML tries to minimize the distances between items and users 
based on the preferences of the users.
Intuitively, it ``pulls'' (decreases the distance of) similar pairs (item, user) closer together in the joint user-item space. During the learning process, users who co-liked the same items will become closer neighbors in the learned vector space. The items which are co-liked by the same users will become close neighbors in the learned low-dimensional space as well. Additional features can also be taken into consideration to bring items with similar tags or features.

CF has been used in recommender systems for a long time~\cite{sarwar2001item, goldberg1992using}. The most common way to model user preferences is to consider them as positive or unknown/negative, commonly known as One-Class Collaborative Filtering~\cite{pan2008one}. Recently, it was shown that modeling user preferences using Two-Class Collaborative Filtering (TCCF) can generate better and diverse recommendations~\cite{paudel2017fewer}. In this work, new approaches based on matrix factorization were proposed to deal with negative feedback.

Signed edges in social networks have also been studied in the machine learning community in the context of link prediction.
%\vspace{-2mm}
\section{Two-Class CML}
In this section, we describe our recommendation algorithm, Two-Class Collaborative Metric Learning (TC-CML).

Let $r_{ij}$ denote user $i$'s rating on item $j$.
The goal is to learn the user vector $u_i \in \mathbb{R}^d$ and item vector $v_j \in \mathbb{R}^d$ in a $d$-dimensional vector space such that the dot product $u_i^T v_j$ approximates $r_{ij}$. The user and item vectors are learned in a way that their Euclidean distance $d(i,j)$ 
%= (u_i - v_j)^2$ 
obeys user $i$'s relative preferences. 
The resulting effect is that items liked by a user will become closer neighbors in the learned vector space, compared to the items which s/he did not interact with.
 
In CML~\cite{hsieh2017collaborative}, a metric is learned using the following loss function:
\begin{align}
\label{cml_opt}
    L_{0} = \sum\nolimits_{(u,i_j) \in S} \sum\nolimits_{(u,i_k)\notin S}w_{u, i_j}[m + d(u,i_j) - d(u,i_k)]_+
\end{align}
where $S$ is the set of observed user-item interaction pairs, $i_j$ is an item liked by user $u$, $i_k$ is an item not interacted by the user, $[z]+ = max(z, 0)$ denotes
the standard hinge loss, and $w_{u,i_j}$ is a ranking loss weight.

While CML only considers positive feedback, our model TC-CML also explicitly includes negative feedback given by the user. From~\eqref{cml_opt}, we see that the CML tries to increase the distance with items that the user did not interact with, while decreasing the distance with items that the users interacted with. It treats all observed user-interactions as positive and all unobserved ones as negative.
In other words, CML treats all user interactions as positive, and ``pulls'' those items closer to the user.

\begin{figure}
    \centering
    \includegraphics[scale=0.5]{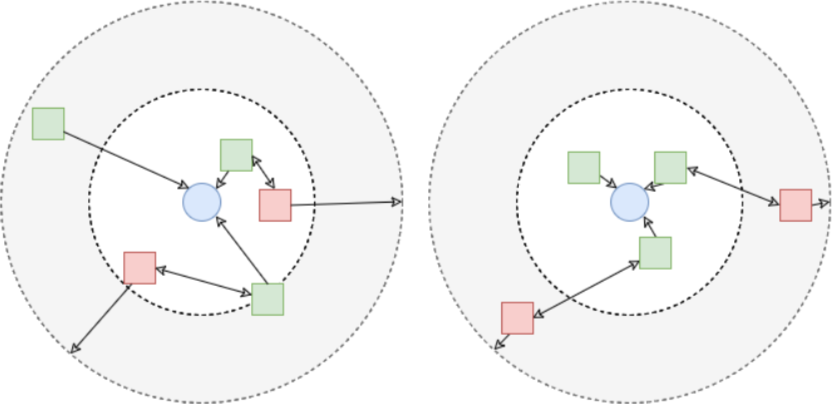}
    \caption{\footnotesize{CML (left) pulls items with observed interactions closer to the user; TC-CML (right) additionally distinguishes observed negative items and pushes them farther. Green and red squares indicate positive and negative items respectively, circle indicates the user.}}
    \label{fig:tc_cml_fig}
    %\vspace{-2mm}
\end{figure}

In contrast, TC-CML distinguishes observed user-interactions into positive and negative classes. It ``pushes'' (increases the distance with) negative items further away from the user, in addition to pulling positive items and pushing unobserved items. This distinction is the reason behind the name Two-Class CML, and is visualized in Figure~\ref{fig:tc_cml_fig}. In Figure~\ref{fig:tc_cml_fig}, the circle denotes a user. Green and red squares indicate positive and negative items, respectively. The effect of CML is shown on the left-hand side of the figure, and that of TC-CML is on the right-hand side. We see that in case of CML, all observed interactions are treated as positive (green squares) and all unobserved interactions as negative.  The effect of CML is to pull items that the user interacted with, and to push all other items. In case of TC-CML, observed interactions are separated into positive (green squares) and negative (red squares) classes. We do not show unobserved items in the illustration for TC-CML. The effect of TC-CML is then to pull positive items closer to the user, and to push negative items away, in addition to pushing the items with unobserved interactions.

To achieve the distinction between positive, negative, and unobserved interactions, we introduce two additional loss terms to to~\eqref{cml_opt}, as given below.
\begin{align}
\label{tccml_opt1}
    L_{1} &= \sum\nolimits_{u\in U}\sum\nolimits_{i \in N_u} d_N^u(u,i)\\
\label{tccml_opt2}
    L_{2} &= \sum\nolimits_{u\in U}\sum\nolimits_{(i,j)\in DP_u} d_{DP}^i(i,j)
\end{align}

In~\eqref{tccml_opt1}, $N_u$ is the set of items rated negatively by the user and $d_N^u(u,i)$ refers to a negative distance measure between user $u$ and item $i$ that s/he rated negatively: $d_N^u(u, i) = max(\frac{\alpha}{d(u, i)}, 1.0)$. 
In ~\eqref{tccml_opt2}, $DP_u$ is the set of dissimilar pairs of items $(i,j)$ such that $i$ received positive while $j$ received negative feedback from the user. The negative distance measure for dissimilar pair of items is $d_{DP}^i(i,j) = max(\frac{\alpha}{d(i, j)}, 1.0)$. 
Both $d_N^u$ and $d_{DP}^i$ increase as the distance $d(.,.)$ between user-item or item-item pair decreases. For negative user-item or item-item pairs, this has the effect of pushing them away. The tunable parameter $\alpha \geq 1$ can be used to set the maximum distance by which such negative items should be pushed away.

\paragraph{\textbf{Optimization and Training.}}

The complete objective function of TC-CML is to minimize the linear combination of loss terms in~\eqref{cml_opt},~\eqref{tccml_opt1}, and~\eqref{tccml_opt2} as given below:
\begin{align}
    \label{tccmf_obj}
    \min_{u*, v*} L_{0} + \lambda_1L_{1} + \lambda_2L_{2} + \lambda_fL_f + \lambda_cL_c
\end{align}
where $\lambda$'s are hyper-parameters to control the weight of loss terms, $L_f$ are additional item-features (e.g., tags) that can be used if available, and $L_c$ describes covariance loss that can also be optionally used~\footnote{We did not use it, as the baseline CML implementation does not use it.}.

The training is done iteratively until convergence or a fixed number of steps using minibatch stochatic gradient descent with Adam optimizer. At each training iteration, we: (i) sample N pairs $(u,i_j)$ of positive user preferences, (ii) sample N pairs of negative user preferences $(u,i_k)$, (iii) sample N pairs from $DP_u$, (iv) for each positive pair, sample U negative items, (v) for each positive pair, keep the negative item k that maximizes the hinge loss and form a mini-batch of size N, (vi) compute gradients and update parameters.

\section{Experiments and Results}
In this section, we describe our experimental setup and present the evaluation of our method in comparison to CML.

We used Tensorflow to implement our model and used the original implementation for CML~\footnote{\url{https://github.com/changun/CollMetric}}. We used two common benchmark datasets from recommender systems: Goodbooks-10k (Goodbooks)~\footnote{\url{https://github.com/zygmuntz/goodbooks-10k}}, with about six-million ratings on books and Movielens 1M (ML-1M)~\footnote{\url{https://grouplens.org/datasets/movielens/1m/}}, with one-million ratings on movies. 
Both datasets contain ratings on 1-5 scale by users on items.
Table~\ref{tab:datasets} shows the properties of the datasets.

For each user, similar to~\cite{paudel2017fewer}, we calculated the mean rating and treated ratings smaller than mean as negative and those equal or greater than the mean as positive.
Both ML-1M and Goodbooks datasets have user-generated tags as item features. 
For ML-1M, we added tags for movies using the \emph{themoviedb.org} API to increase their number.
These tags were used as item features for both ML-1M and Goodbooks. To exclude spelling errors and rare entries, we removed tags that were given by less than five users or those used only for a single item.

We randomly divided the datasets into three equal parts for training, validation and testing. From the test-set, users with at least three positive feedbacks were used for evaluation. 
Hyper-parameters were chosen using grid-search.
%on a limited set of choices.
Embedding dimension $d=70$ was used for ML-1M and $d=100$ for Goodbooks.
We ran each evaluation for three times and report the average results.

To measure the quality of recommendations, we used the following measures: Precision, Recall, and Negative-Items at top-k (NI@k). Precision (P@10, P@50) and Recall (R@10, R@50) measure the recommendation accuracy. To measure how well our approach is able to remove negative items from the top of recommendations, we adopt the NI@k (NI@10, NI@50) measure from~\cite{paudel2017fewer}, which calculates the fraction of negatively rated items at the top-k of the recommendation list. 
Higher numbers for Precision and Recall indicate more accurate recommendations, and lower numbers for NI@k indicate fewer negative recommendations.

\begin{table}[]
    \centering
    \begin{tabular}{c|c|c|c|c|c}
        \hline
        Dataset & \#Users & \#Items & \#Ratings & Density & Avg Rating \\
        \hline
        Goodbooks & 53,424 & 10,000 & 5,976,479 & 0.011 & 3.92 \\
        \hline
        ML-1M & 6,040 & 3,952 & 1,000,000 & 0.042 & 3.58 \\
        \hline
    \end{tabular}
    \caption{Description of the datasets, with \#users, \#items, \#ratings, density, and average rating.
    }
    \label{tab:datasets}
%\vspace{-3mm}
\end{table}

\begin{table}[h]
    \centering
    \begin{tabular}{c|c|c|c|c|c|c}
        \hline
         Method & R@10 & R@50 & P@10 & P@50 & NI@10 & NI@50 \\
         \hline \hline
         CML & 0.82 & 0.24 & 0.56 & 0.76 & 0.020 & 0.05 \\
         \hline
         TC-CML & \textbf{0.92} & \textbf{0.26} & \textbf{0.61} & \textbf{0.79} & \textbf{0.018} & \textbf{0.04} \\
         \hline
    \end{tabular}
    \caption{Experimental evaluation on the Goodbooks dataset.}
    \label{tab:goodbooks}
%\vspace{-3mm}
\end{table}

\begin{table}[h]
    \centering
    \begin{tabular}{c|c|c|c|c|c|c}
         \hline
         Method & R@10 & R@50 & P@10 &P@50 & NI@10 & NI@50 \\
         \hline \hline
         CML & 0.12 & 0.32 & 0.63 & \textbf{0.74} & 0.031 & \textbf{0.12} \\
         \hline
         TC-CML & \bf{0.13} & 0.33 & \bf{0.66} & \textbf{0.74} & \bf{0.028} & \textbf{0.12} \\
         \hline
    \end{tabular}
    \caption{Experimental evaluation on the Movielens dataset.}
    \label{tab:ml1m}
    %\vspace{-5mm}
\end{table}

Experiments comparing CML and TC-CML for Goodbooks and ML-1M are summarized in Table~\ref{tab:goodbooks} and Table~\ref{tab:ml1m} respectively. Better results are boldfaced. For Goodbooks, we can see that TC-CML outperforms CML in all measures. 
In the denser ML-1M dataset, the improvements of TC-CML are more pronounced at the very-top (top-10) of the recommendations lists, where it outperforms CML in all measures.

From these experiments, we see that TC-CML produces recommendations that are both more accurate and have fewer negative items in the recommendations. The results show the benefit of modeling the context provided by negative preferences in TC-CML.
\section{Conclusion and Future Work}
We described TC-CML, a new recommendation algorithm which models negative user preference using a metric learning framework.
With experimental results, we showed that TC-CML 
is able to improve recommendation quality on two benchmark datasets, demonstrating the benefit of utilizing the context provided by negative user-feedback. 
In future, we would like to compare TC-CML with other baseline algorithms. 
We would also like to evaluate the performance of TC-CML on more datasets and applications, including social-network link prediction.
We would also like to consider other aspects like diversity and novelty of recommendations.
In this work, we explored one particular method to convert implicit user-feedback into positive and negative classes. In the future we would like to compare our algorithm on explicit feedback datasets, and using other ways to classify implicit user-feedback.
\newpage

\bibliographystyle{ACM-Reference-Format}
\bibliography{07_bibliography}

\end{document}